\documentstyle [prl,aps]{revtex}
\input epsf
\begin{document}

\draft
\title{Development of Bubble Chambers With Enhanced Stability and  
Sensitivity to Low-Energy Nuclear Recoils}

\author{
W.J.Bolte$^{a}$,
J.I.Collar$^{a,*}$, M.Crisler$^{b}$, J.Hall$^{a}$, D.Holmgren$^{b}$, 
D.Nakazawa$^{a}$, B.Odom$^{a}$,
K.O'Sullivan$^{a}$, R.Plunkett$^{b}$, 
E.Ramberg$^{b}$, 
A.Raskin$^{a}$, A.Sonnenschein$^{a}$ and J.D.Vieira$^{a}$
}
\address{ 
$^{a}$Enrico Fermi Institute and Kavli Institute for Cosmological 
Physics, 
University of Chicago,
IL, USA\\
$^{b}$Fermi National Accelerator Laboratory, Batavia, IL, USA\\
}
%\date{\today}
\wideabs{
\maketitle
%\tightenlines
\begin{abstract}
\widetext
The viability of using a Bubble Chamber for rare event searches 
and in particular for the detection of dark matter particle candidates 
is considered. Techniques leading to the deactivation 
of inhomogeneous nucleation 
centers and subsequent enhanced stability in 
such a detector
are described. Results from prototype trials indicate 
that sensitivity to low-energy nuclear recoils like those expected 
from Weakly Interacting Massive Particles can be obtained in 
conditions of near total insensitivity to minimum ionizing backgrounds. An 
understanding of the response of superheated heavy refrigerants 
to these recoils is demonstrated within the context of existing theoretical 
models. We comment on the prospects for the detection of supersymmetric dark matter 
particles with a large $CF_{3}I$ chamber. 
\end{abstract}

%insert suggested PACS numbers in braces on next line
\pacs{PACS number(s): 
95.35.+d, 29.40.-n, 05.70.Fh, 68.55.Ac\\
$^{*}$~Corresponding author. E-mail: collar@uchicago.edu}}
\narrowtext

The positive identification of sporadic signals from among comparatively 
frequent backgrounds is common to any experiment at the 
forefront of particle physics. The challenge 
faced by direct searches for cold dark matter particles \cite{revexp}
is in this respect extraordinary: signal rates as small as 
one low-energy nuclear recoil (few keV) per ton of detector mass per year are 
predicted for the  nuclear scattering of 
supersymmetric Weakly Interacting Massive Particle (WIMP)
candidates, if they comprise 
the bulk of dark matter halos able to explain galactic evolution and 
dynamics \cite{revtheo}. A number of detector techniques have been 
developed during the last two decades for this purpose 
\cite{revexp}. Simplicity of design, optimal target 
materials, rapid scaling to the ton regime 
and an excellent background 
rejection are desirable qualities for the next-generation of
detectors that 
should soon explore the 
vast range of WIMP masses and couplings still allowed. 

The use of moderately
superheated liquids has been proposed as a possible fast route towards this 
goal \cite{zacek,myprd}. A concentrated energy deposition from 
certain particles can lead in these 
to the rupture of metastability and the formation of visible bubbles. Two 
experiments, SIMPLE \cite{prlsdd} and PICASSO \cite{recpic} exploit 
this approach, benefiting from an intrinsic insensitivity to most 
backgrounds, discussed below. Both experiments implement the method 
using superheated droplet detectors 
\cite{apfel1} (SDDs, 
a.k.a. bubble detectors), where small drops (r $\sim\!10\mu m$) 
of the active liquid are dispersed in an insoluble
gel or viscoelastic medium. 
In a SDD the gel provides a smooth liquid-liquid interface that
impedes the continuous triggering (inhomogeneous nucleations) 
on surface defects, gaskets, motes, etc.
that is observed 
even in the cleanest bubble chambers. As a result, the lifetime 
of the superheated state is
considerably extended, to the point that a WIMP search can be 
performed. 

The goal of the present study is to assess the feasibility 
of employing bulk quantities of superheated liquid 
instead, i.e., to use a conventional bubble chamber, an alternative for WIMP 
searches first put 
forward by Hahn \cite{hahn}. Large, stable bubble chambers have been 
previously proposed for other rare-event searches (e.g., nucleon 
decay, superheavy elements
\cite{protond}) but no dedicated attempt to extend the 
superheated times
was made. The rapid uncontrollable foaming of a conventional 
chamber following its decompression was bypassed in 
accelerator experiments by a precise timing of the pulsed beam injection 
to coincide with the few ms of usable radiation-sensitive superheated time 
in each pressure cycle. The 
motivation to explore this apparently more problematic approach
arises from the difficulty to manufacture SDDs out of the most 
interesting available industrial refrigerants, 
e.g., $CF_{3}I$ 
and $CF_{3}Br$. These liquids constitute ideal supersymmetric 
WIMP targets \cite{mucf3i} due to the presence of both fluorine (optimal for 
spin-dependent neutralino couplings \cite{john}) and a heavy nucleus (maximally sensitive to 
coherent spin-independent couplings) \cite{njp,othersd}. Their 
density is nevertheless severely mismatched with respect to that of 
a water-based SDD gel matrix, leading to inhomogeneous, unstable 
emulsions during the fabrication process. 
Saturation of the matrix with inorganic salts can help alleviate this 
issue, but leads to exacting requirements on the alpha-emitter 
radiopurity of the gel \cite{recpic}, exacerbated by the observed 
tendency of complex actinide 
salts to migrate to the droplet-gel boundary \cite{njp,wang}, where their ability to 
create an undesirable alpha-recoil background is the greatest. A first 
attempt to measure the attainable stability of bulk 
superheated liquid was made within the context 
of the SIMPLE experiment, using a rudimentary 
plastic chamber where the active fluid was fully encapsulated by a
thick sheath of viscoelastic liquid \cite{aqua} 
to avoid evaporation and nucleations on chamber walls. 
The chamber held 30 g of R-115 ($C_{2}ClF_{5}$) superheated 
for up to 12 hours at an underground depth of 1,500 
m.w.e. \cite{joel}, with 
no other precautions against neutron or radon  backgrounds. 
This behavior revealed the possibility to control the sources of instability 
in a bubble chamber and prompted the further experimentation described 
here.
\begin{figure}[tbp]
\epsfxsize = \hsize \epsfbox{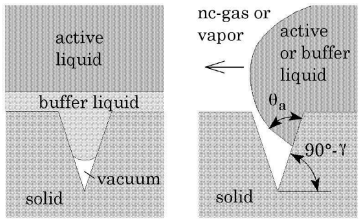}
\caption{{\it Left}: Use of a buffer liquid to isolate microscopic surface cavities 
able to act as inhomogeneous nucleation centers in a bubble chamber 
\protect \cite{reinke}. Mass 
transfer into the cavity can still lead to boiling, but deactivation is possible in the absence 
of noncondensable (nc) gas (see text). {\it Right}: Direct pouring of a liquid 
during chamber filling can lead 
to vapor entrapment in cavities when the advancing contact angle 
$\theta_{a}$ is larger than the groove angle $2\gamma$ \protect 
\cite{book}. 
Filling by slow vapor condensation after evacuation instead leads to 
efficient wetting 
of cavities, including those reentrant.} 
\end{figure}

The mechanism leading to the nucleation of the gaseous phase and possible 
ensuing phase transition in the bulk of a superheated liquid 
(homogeneous nucleation) is described by a
classical theory \cite{homog} where the probability of 
spontaneously generating a protobubble of radius larger than $r_{c}$ 
is computed as a function of pressure, temperature and 
thermophysical properties of the liquid. If this critical radius 
is reached or surpassed, the vapor 
nucleus grows unchecked and metastability is lost. 
Protobubbles smaller than $r_{c}$ collapse back
onto themselves, producing no phase transition.
In the case of a radiation-induced nucleation, the 
local heating (``hot spike'' \cite{seitz}) from a particle's energy 
deposition can be responsible for the formation of a critically-sized 
nucleus, only if this 
energy is concentrated over a small enough region, comparable to 
$r_{c}$. This leads to the mentioned insensitivity to most backgrounds 
by imposing a threshold in stopping power, amply surpassed by nuclear 
recoils but not by minimum-ionizing particles (MIPs) \cite{zacek,myprd}.
For moderate degrees 
of superheat like those necessary to sensitize the liquid to 
low-energy nuclear 
recoils, $r_{c}$ is
typically few tens of nm, and the rate of homogeneous nucleation 
is entirely negligible as a source of instability 
($<< 10^{-20}$ bubbles/kg/day \cite{homog}).

Nucleations can nevertheless also occur on microcavities, scratches
or imperfections naturally 
present even in the smoothest surfaces (e.g., glass) or in  
motes, partially or totally wetted by the liquid. 
The (inhomogeneous) nucleation rate on these cavities is only 
minimally increased with respect to the extremely small (homogeneous)
bulk rate 
if the fluid has a zero contact angle with the surface, i.e., if the 
cavity is well wetted \cite{homog,book}. The actual source of the inhomogeneous 
nucleations that limit a bubble chamber's stability
is instead any entrapped gas in these cavities\cite{nature}, which 
can 
act as a vaporization initiator, allowing mass transfer from the fluid 
to the unwetted  cavity  volume\cite{book,reinke}. Once 
nucleation is initiated in such a cavity, the superheat required to 
sustain boiling on it drops to a much lower value than what is 
required for homogeneous nucleation\cite{book,reinke}, i.e., destabilization occurs. It is 
however important to distinguish between cavities filled by 
superheated fluid 
vapor and those filled by noncondensable gas or a binary. In the first case, cooling 
or pressurization can lead to nucleation site deactivation
by recondensing the trapped vapor. In the second, and 
in particular for reentrant cavities, deactivation can be 
arduous, albeit continued boiling may lead to an eventual depletion of 
the gaseous volume\cite{book}.

Once the nature of the problem is understood, precautions can be taken 
that lead to an enhanced bubble chamber stability:
{\it i}) only smooth glass or quartz surfaces are allowed to be 
wetted by the superheated liquid, thereby reducing the number 
of available cavities. {\it ii}) A layer of a low-density 
buffer liquid can be allowed to form a ``lid'' above the (immiscible)
active 
liquid \cite{lid}, with all rough metallic parts (bellows, diaphragms, 
gaskets) coming in 
contact with the buffer only. {\it iii}) This same buffer liquid can 
be used to create a layer that fills cavities, previously evacuated
to remove noncondensable gases
({\frenchspacing Fig. 1, left})\cite{reinke}. Cavity filling 
can be improved by 
transferring the buffer (a step prior to the addition of the denser active 
liquid) by slow condensation of its vapor into the chamber rather 
than pouring. 
This ensures maximum wetting of even reentrant cavities 
\cite{book,private} ({\frenchspacing Fig. 1, right}). In the particular case of $CF_{3}I$, 
the shape of the meniscus at the interface with the
buffer ``lid'' reveals a highly preferential wetting of  
quartz by the (water) buffer, a positive indication of its 
effectiveness. To some extent, these methods reproduce the 
advantages of the
smooth liquid-liquid interfaces in SDDs. {\it iv}) Exhaustive cleaning 
of glass surfaces \cite{clean1} in clean-room conditions and 
ultrafiltration of all gases and fluids lead to a reduction in the number of 
large motes 
present (cavities smaller than O($r_{c}$)  cannot in principle act as 
nucleation centers). Some known cleaning techniques also have the 
desirable effect of improving surface wetting by the buffer \cite{clean2}. 
{\it v}) After application
of these techniques in the chambers and operating 
conditions described below, a periodic 
long recompression ($\sim$ 200 s) is observed to 
effectively deactivate the few boiling centers that can still sporadically appear 
due to mass transfer through the buffer layer, or cavity exposure  
to vapor during radiation-induced boiling. 

Small bubble chamber prototypes up to 50 {\frenchspacing 
c.c.} in active volume
can 
be built for moderately-superheated refrigerants 
using commercially available pressure-resistant quartz vials 
\cite{griff}. 
Pressure cycling is achieved with a 
three-way valve or its equivalent and temperature control by means 
of a double-bath \cite{taup}. Fast triggering ($<$10 ms) 
of bubble photography and recompression can be performed by
use of a piezoelectric microphone to detect the 
acoustic emission  
that accompanies nucleations \cite{njp} or by 
monitoring the  
pressure increase caused by bubble growth. These simple devices have 
been 
used to study chamber stability and response to radiation sources.
\begin{figure}[tbp]
\epsfxsize = \hsize \epsfbox{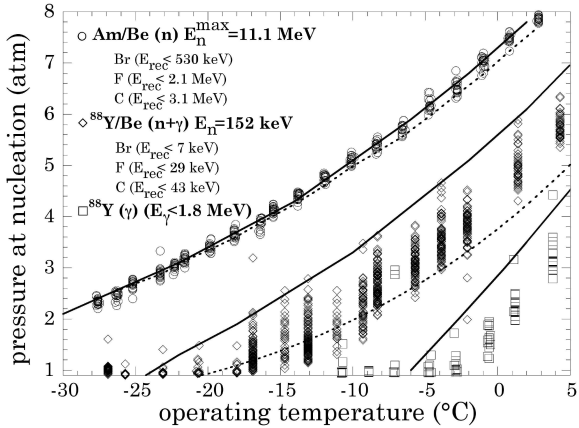}
\caption{Response of a $CF_{3}Br$ chamber to radiation sources and comparison 
with theoretical models. 
Lines indicate the pressure below 
which full sensitivity to the source is expected according to 
different  
theoretical models (the experimental points represent the 
appearance of the first bubble upon decompression, see text).
Insensitivity to gamma interactions in 
conditions that nevertheless afford good sensitivity to 
low-energy nuclear recoils has been demonstrated (see text).} 
\end{figure}
Calibrations using neutron sources having a well-defined maximum 
energy (11.1 MeV for $^{241}$Am/Be) or monochromatic neutron 
emission (152 keV for $^{88}$Y/Be) 
have allowed measurement of the response of the liquids to nuclear 
recoils down to 4 keV in the case of $CF_{3}I$ and to establish agreement with 
theoretical models of this response. 
Data points in {\frenchspacing Fig. 2} represent the appearance of the first bubble nucleation 
upon decompression in the presence of each source (i.e., as the energy 
threshold for radiation-induced nucleation is reduced), each point corresponding 
to a compression/decompression cycle. For sufficiently-high source 
intensities and/or slow decompression rates this bubble is 
the result of a recoil with an energy close to the well-defined 
maximum 
that these sources can 
produce. These maximum recoil energies are indicated by labels in the 
figure, for 
each recoiling species. 
Solid lines represent the 
theoretical expectations (Seitz model \cite{seitz,harper})
for the onset of sensitivity to maximum-energy recoils,
i.e., should trace the top boundary of 
the data points. Their dispersion towards lower pressures 
is expected from a progressive onset of sensitivity, which is not 
well described by a 
step-function \cite{recpic} as naively assumed in the Seitz model. 
The effect of  
Moliere electron
straggling \cite{archard} was
included in the calculation of  $^{88}Y$ (gamma) response.
A review of the theoretical background leading to 
these predictions can be found in \cite{njp}.
A good agreement with the data is observed by 
best-fitting the single free parameter in this model. 
The best value 
obtained ($a\!\sim$4 in the notation of \cite{njp}) is compatible with 
previous \cite{harper} and most recent \cite{das}  
studies. 
Since the predicted onset of response to the source for each recoiling species 
is not exactly the same (differing by a fraction of an atm), 
the lines represent the first species expected to react to 
the source (Br and F, closely matched, for $CF_{3}Br$). 
A calibration is planned where tagging of gamma rays emitted 
in 
inelastic 2.4 MeV neutron scattering will allow to separate the  
contributions from each species. Dotted lines correspond to the 
$^{241}$Am/Be and $^{88}$Y/Be predictions of a modern phenomenological ``reduced superheat'' theory, 
the forte of which is its simplicity. It generates remarkably accurate 
predictions for lighter refrigerants 
used in SDDs \cite{rs}, 
but appears to need further refinement for $CF_{3}Br$ and $CF_{3}I$ \cite{rs2}.

The photonuclear $^{88}Y/Be$ source employed emitted a mixed field of 
$\sim\!\!10^{8}$ 
high-energy gammas and just $3.5\times10^{3}$ monochromatic neutrons per 
second: {\frenchspacing Fig. 2} illustrates the much higher degree of superheat (lower 
pressure at a given temperature) necessary to become 
sensitive to the gamma component once the Be sheath, the actual neutron 
emitter, is removed from 
the source.
This allows for a dramatic demonstration of insensitivity to 
photoelectrons in operating conditions that nevertheless would
ensure an optimal response to WIMP 
interactions. For instance, from the figure, at -10$^{\circ}$C and 1 atm no 
response to MIPs is observed, while sensitivity to WIMP-induced recoils
more energetic than the maximum recoil energies produced by $^{88}Y/Be$ seems 
guaranteed. A recently procured $^{124}Sb/Be$ source 
($E_{n}$= 24 keV)
will be used to calibrate response to
recoil energies $\sim 1$ keV in $CF_{3}I$, an unprecedented test 
of a WIMP detector. 

Prototypes containing a few tens of {\frenchspacing c.c.} of active liquid 
remain superheated for periods of several 
minutes on the 
average in a shallow-depth laboratory (6 {\frenchspacing m.w.e.}). 
The reduced ambient neutron flux in this site was characterized 
using a $^{3}$He detector surrounded by several configurations of 
neutron moderator and absorber (Bonner spheres) 
calibrated using known neutron sources, 
and deconvolved following an approach similar to \cite{belli}. 
Taking the measured fast neutron 
spectrum as input to a MCNP-POLIMI simulation \cite{polimi} of the 
energy 
depositions in the chamber, the observed spontaneous
nucleation rate is found to be in agreement with the expected neutron-induced recoils at 
this depth ({\frenchspacing Fig. 
3}). For superheated times $t_{SH}$ longer than a few seconds, 
no observable excess of nucleations on walls can be inferred from 
bubble photography using two orthogonal cameras, which allows for 3-D
reconstruction of nucleation sites with $\sim$1 mm precision. For 
shorter $t_{SH}$ a small excess of wall events, evident in the figure, 
is observed. Sporadic boiling sites can be deactivated as previously described. 
The duty (live) time was $\sim$65\% during these runs. The insensitivity (rejection 
factor)
to MIPs 
at $-10^{\circ}$C and
1 atm is $\gtrsim 10^{9}$ from the 
absence of any observable reaction to the $\sim\!10^{6}$ gamma interactions per second 
induced by the $^{88}Y$ source 
within the active volume ({\frenchspacing Fig. 3}). As discussed above, 
good sensitivity to WIMP-recoils is nevertheless expected in these conditions. This 
{\it intrinsic} rejection factor can be compared with the best 
($\sim\!\!
10^{4}$) achieved 
using complex cryogenic WIMP detectors \cite{revexp}. It should
permit construction of much larger chambers in the 
ton or multi-ton regime essentially without any concern for 
MIPs, including from elevated concentrations of $^{14}$C.
\begin{figure}[tbp]
\epsfxsize = \hsize \epsfbox{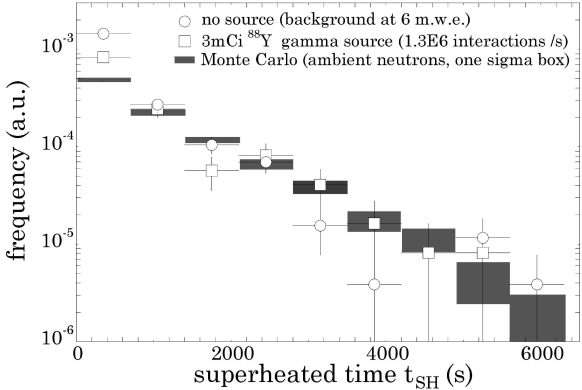}
\caption{Distribution of duration of the superheated state $t_{SH}$
in a 12 ml $CF_{3}Br$ bubble chamber operated at 6 m.w.e., -10$^{\circ}$C and
atmospheric pressure. When fitted by a form $\propto 
e^{-t_{SH}/\tau}$ 
and after rejection of inhomogeneous nucleations during 
decompression, all cases depicted yield $\tau\sim$1220 s.} 
\end{figure}

The encouraging outcome from these tests has lead to the construction of 
a steel recompression chamber housing 2 kg of $CF_{3}I$ in an inner quartz vessel 
\cite{edi}. 
A bellows mechanism compensates the pressure inside and outside of 
this vessel, which is both sealed against Rn penetration and low in 
its emanations, measures against alpha-recoil 
backgrounds. Provisions to ensure the long-term stability of this 
fire-extinguishing compound at the envisioned running temperature 
(40$^{\circ}$C) are in place \cite{nist}. The behavior of this chamber at 6 
m.w.e. ($<\!\!t_{SH}\!\!>\sim$60 s)
remains in agreement with the simulated ambient neutron contribution. Ongoing 
tests of $CF_{3}I$ neutron and gamma response yield similar results to those 
discussed. To 
further assess the prospects of this new approach to WIMP 
detection, this chamber will be operated during 2005 within a neutron shield  
at the 300 m.w.e. 
depth of the MINOS near gallery on FNAL grounds (COUPP, the 
Chicagoland Observatory for Underground Particle Physics). 
The WIMP sensitivity 
that can in principle be achieved with COUPP
is already competitive with 
the best present searches \cite{edi}. Besides those mentioned, there are 
unique advantages (neutron rejection ability, 
rapid target replacement, room 
temperature operation and low cost \cite{edi})
that distinguish ton-sized bubble 
chambers   
from their competitors 
in this exciting endeavor of dark matter detection: a successful 
COUPP would lead us to attempt their construction.

We are indebted to {\frenchspacing F.d'Errico}, {\frenchspacing J.Ely}, 
{\frenchspacing D.Jordan}, 
{\frenchspacing E.Padovani}, {\frenchspacing D.Qu\'{e}r\'{e}}, 
and in particular to {\frenchspacing R.Hildebrand} for being a constant source of inspiration.
\uppercase{W}ork
supported by the \uppercase{K}avli \uppercase{I}nstitute 
for \uppercase{C}osmological \uppercase{P}hysics 
 (\uppercase{N}\uppercase{S}\uppercase{F} 
grant \uppercase{PHYS}-0114422) and \uppercase{N}\uppercase{S}\uppercase{F} 
\uppercase{CAREER} award 0239812.

\end{document}